\documentclass{ptephy_v1}

\usepackage{hyperref}
\usepackage{dcolumn}

\definecolor{emeraldgreen}{rgb}{0.0,.6015625,.40625}

\newcommand{\gsim}{\, \raisebox{-0.8ex}{$\stackrel{\textstyle >}{\sim}$ }}

\begin{document}

\title{New constraints on the neutron-star mass and radius relation from terrestrial nuclear experiments}


\author[1,2,*]{Hajime Sotani}
\author[1,3,4]{Nobuya Nishimura}
\author[5,3]{Tomoya Naito}
\affil[1]{Astrophysical Big Bang Laboratory, RIKEN, Saitama 351-0198, Japan
    \email{sotani@yukawa.kyoto-u.ac.jp}}
\affil[2]{Interdisciplinary Theoretical \& Mathematical Science Program (iTHEMS), RIKEN, Saitama 351-0198, Japan}
\affil[3]{RIKEN Nishina Center for Accelerator-Based Science, Saitama 351-0198, Japan}
\affil[4]{Division of Science, National Astronomical Observatory of Japan, Tokyo 181-8588, Japan}
\affil[5]{Department of Physics, Graduate School of Science, The University of Tokyo,
    Tokyo 113-0033, Japan}


\begin{abstract}%
The determination of the equation of state (EOS) for nuclear matter has been one of the biggest problems in nuclear astrophysics, because the EOS is essential for determining the properties of neutron stars. 
To constrain the density-dependence of the nuclear symmetry energy, several nuclear experiments, e.g., reported by the S$\pi$RIT and PREX-II collaborations, have recently been performed. However, since their uncertainties are still large, additional constraints such as astronomical observations must be crucial.
In addition, it is interesting to see the effect of their reported value on neutron star properties. In this study, focusing on the relatively lower density region, we investigate the allowed area of the neutron-star mass and radius relation by assuming the constraints from S$\pi$RIT and PREX-II. Each region predicted by these experiments is still consistent with the allowed area constrained by the various astronomical observations. Our results show that terrestrial nuclear experiments must provide further constraints on the EOS for neutron stars, complementing astronomical observations.
\end{abstract}

\subjectindex{E32, D41}

\maketitle


\textit{1.~Introduction.}
Core-collapse supernovae, which occur at the end of a massive star's life, produce a neutron star (NS) (or a black hole) as a compact remnant. Since the density inside the NS becomes significantly larger than the nuclear saturation density, which must be the densest environment in nature, the nuclear equation of state (EOS) describing NS matter is not fixed yet. So, astronomical observations help us to understand the properties in such a high-density region. For example, the discovery of NSs heavier than $2M_{\odot}$ could rule out some soft EOSs, with which the theoretical maximum mass does not reach the observed mass \cite{Demorest:2010bx, Antoniadis:2013pzd, NANOGrav:2019jur}. The observation of the gravitational waves from the binary NS merger, GW170817, tells us the constraint on the tidal deformability, which estimates that the $1.4M_\odot$ NS radius should be less than $ 13.6 \, \mathrm{km} $ \cite{Annala:2017llu}. The electromagnetic signals from the NS also give a constraint on the neutrons star mass and radius, although it depends on the theoretical models \cite{Steiner:2012xt,Miller:2019cac,Riley:2019yda}. In this way, the NS observations essentially give us the information and/or constraint on the relatively high-density region. 

As it has been traditionally performed, on the other hand, the terrestrial particle-accelerator experiments are still crucial to obtain information on the physics of nuclear matter. Owing to the nuclear saturation properties, one may easily constrain the EOS around the nuclear saturation density. In fact, several experiments have been performed to determine the nuclear saturation parameters, especially focusing on nuclear symmetry energy.
Here, the symmetry energy is roughly the difference between the energy for symmetric nuclear matter and that for pure neutron matter. This is one of the key properties for constructing the EOS for NS matter because NS matter is very neutron-rich under the $\beta$-equilibrium state. Even so, the access to the symmetry energy via terrestrial experiments is relatively more difficult than that for the other nuclear properties, because the stable atomic nucleus on the Earth is at most $\alpha \simeq 0.3$, where $\alpha$ is an asymmetry parameter. We note that $\alpha = 0$ and $ 1 $ correspond to the symmetric nuclear matter and pure neutron matter, respectively.

Recently, new experiments in two large facilities have reported a constraint on the density-dependence of symmetry energy, $L$ 
(see Eq. (\ref{eq:Sn}) for definition); 
$42\le L \le117 \, \mathrm{MeV}$ by the Radioactive Isotope Beam Factory (RIBF) at RIKEN in Japan (S$\pi$RIT, e.g., \cite{SRIT:2021gcy}) and $L = 106 \pm 37\, \mathrm{MeV} $ obtained by the polarized-electron scattering done at the Thomas Jefferson National Accelerator Facility in Newport News, Virginia, the United States (PREX-II,  \cite{PhysRevLett.126.172502,Reed:2021nqk}). 
Compared to the previous $L$ values, the PREX-II result suggests a significantly large $L$, while the S$\pi$RIT result also supports a relatively large value. On the other hand, the constraint through the polarized-proton scattering experiment performed at Research Center for Nuclear Physics (RCNP), Osaka University, Japan \cite{Zenihiro:2010zz} is more or less consistent with the other predictions obtained so far, even though the same nuclear property, i.e., the neutron-skin thickness, has been measured at RNCP and PREX-II, but via a different probe. 
So, current terrestrial experiments cannot solely determine the value of $L$, where astronomical observations must be important, which is a qualitatively different approach and covers higher nuclear densities.
In the context of astrophysics, constraints on $L$ become more informative if they are interpreted on the NS mass and radius relation, which can be easily compared with other astronomical constraints by the X-ray and the gravitational wave observations.

In this study, therefore, we concretely show the 
allowed region in the NS mass and radius relation, based on the two new constraints on $L$ (S$\pi$RIT and PREX-II) and another experimental constraint previously obtained (RCNP). The NS mass and radius curves theoretically constructed with the EOSs should be compared with these constraints, as well as the several astronomical restrictions given by X-ray and gravitational-wave observations. We discuss the consistency in the constraints between the nuclear experiments and astronomical observations.



\textit{2.~EOS and nuclear saturation parameters.}
In order to construct the NS models, one has to prepare the EOS for NS matter. For any nuclear EOSs, the bulk energy per nucleon for nuclear matter with zero temperature, $w$, is written as a function of the baryon number density $n_{\text{b}}$ and the asymmetry parameter $\alpha$, as
\begin{equation}
  w(n_{\text{b}},\alpha) = w_{\text{s}} (n_{\text{b}}) + \alpha^2 S(n_{\text{b}}) + \cdots, \label{eq:wn}
\end{equation}
where $n_{\rm b}$ and $\alpha$ are defined by $n_{\text{b}}=n_n + n_p$ and $\alpha=(n_n - n_p)/n_{\text{b}}$ with the neutron number density $n_n$ and the proton number density $n_p$.
In this expansion, $w_{\text{s}}$ and $S$ correspond to the energy per nucleon of symmetric nuclear matter, i.e., $w_{\text{s}} = w(n_{\text{b}},0)$, and the density-dependent symmetry energy given by $S(n_{\text{b}})= \partial w/\partial \alpha^2|_{\alpha=0}$, respectively. Additionally, $w_{\text{s}}$ and $S$ can be expanded as a function of $u\equiv (n_{\text{b}} - n_0)/(3n_0)$ with the saturation density of symmetric nuclear matter, $n_0$, as
\begin{align}
  w_{\text{s}} (n_{\text{b}}) &= w_0 + \frac{K_0}{2}u^2 + \frac{Q_0}{6}u^3 + \cdots, \label{eq:ws} \\
  S(n_{\text{b}}) &= S_0 + Lu + \frac{K_{\text{sym}}}{2}u^2 + \frac{Q_{\text{sym}}}{6}u^3 + \cdots. \label{eq:Sn}
\end{align}
The coefficients are referred to as the nuclear saturation parameters. In particular, among these parameters, the saturation parameters in the lowest order, such as $n_0$, $w_0$, $K_0$, $S_0$, and $L$, are the most important, which strongly associated with the properties of the atomic nuclei in nature, and are constrained via terrestrial experiments. Even so, $n_0$, $w_0$, and $S_0$ are relatively well-constrained, while $K_0$ and $L$ are more difficult to be constrained. This is because one can easily obtain the nuclear information around the saturation density, owing to the nuclear saturation properties,
whereas, to obtain parameters associated with the density derivative, such as $K_0$ and $L$, 
information of nuclear matter properties at various densities is needed to be measured.

Thus, in this study, we focus on $K_0$ and $L$, where $n_0$, $w_0$, and $S_0$ must be tuned in such a way that the properties of stable nuclei should be reproduced by any EOSs. Via the data for the isoscalar giant monopole resonance in $^{208} \mathrm{Pb} $ and $^{90} \mathrm{Zr}$, $K_0$ is constrained in the range of $K_0=240\pm20 \, \mathrm{MeV} $ \cite{Shlomo:2006}, which seems to be conservative constraint on $K_0$ \cite{Garg:2018uam}. On the other hand, there are several attempts for constraining $L$, which tells us that $L$ should be in the range of $L\simeq 60\pm 20 \, \mathrm{MeV}$ \cite{Vinas:2013hua,Li:2019xxz}. Nevertheless, the recent experiments seem to predict a larger value of $L$ than the previous constraints, as mentioned in the next section.


\textit{3.~Experimental constraints.}
In this study, we especially focus on two recent terrestrial nuclear experiments, i.e., S$\pi$RIT and PREX-II, together with RCNP. S$\pi$RIT is the experiment with the isotope beams provided by the RIBF at RIKEN in Japan, where the beams of $^{132}$Sn, $^{124}$Sn, $^{112}$Sn, and $^{108}$Sn were bombarded to the targets of $^{124}$Sn and $^{112}$Sn.
Throughout such reactions, $ \Delta $ isobars were produced,
which decay to nucleons with emitting pions.
The ratio of the production rate of positive charged pions, $ \pi^{+} $, to that of negative one, $ \pi^{-} $, allows one to constrain $ L $ as $42\le L \le117\,\mathrm{MeV}$ with the $1\sigma$ accuracy  \cite{SRIT:2021gcy}.
PREX-II is the experiment after PREX ($^{208}$Pb Radius Experiment) at the Thomas Jefferson National Accelerator Facility in Virginia.
In PREX/PREX-II experiments, first, the scattering cross section of spin-up polarized electrons $ \sigma_{\uparrow} $ and that of spin-down ones $ \sigma_{\downarrow} $ in $^{208}$Pb have been measured. Then, using the parity-violating asymmetry 
$ \left( \sigma_{\uparrow} - \sigma_{\downarrow} \right) / \left( \sigma_{\uparrow} + \sigma_{\downarrow} \right) $,
one can estimate the neutron root-mean-square radius, and accordingly, the neutron-skin thickness,
whose values are $\Delta r_{np} = 0.33^{+0.16}_{-0.18} \, \mathrm{fm}$ in PREX \cite{PhysRevLett.108.112502} and $\Delta r_{np} = 0.283\pm 0.071 \, \mathrm{fm}$ in PREX-II \cite{PhysRevLett.126.172502}. 
By using the data for neutron-skin thickness, the value of $L$ is constrained to be $L = 106 \pm 37$ MeV in PREX-II with the $1\sigma$ accuracy \cite{Reed:2021nqk}. 
We remark that, soon after the report of PREX-II, 
the reanalysis has been done, using the same data for the parity-violating asymmetry in PREX-II, 
which predicts $\Delta r_{np} = 0.19\pm 0.02$ fm and $L$ becomes $L=54\pm 8$ MeV \cite{PhysRevLett.127.232501}. 
\par
In addition to the two above constraints on $L$ derived by S$\pi$RIT and PREX-II, we also consider the experiment at the RCNP, where the neutron density distributions of $^{204, 206, 208}$Pb were measured via polarized-proton elastic scattering and the neutron-skin thickness
especially for $^{208}$Pb was
deduced to be $\Delta r_{np} = 0.211^{+0.054}_{-0.063}$ fm \cite{Zenihiro:2010zz}. It is theoretically known that the neutron-skin thickness is strongly correlated with the slope parameter $L$ \cite{Roca-Maza:2011qcr}, as
\begin{equation}
  \Delta r_{np}\ ({\rm fm}) = 0.101 + 0.00147 L, \label{eq:fit}
\end{equation}
where $L$ is considered in the unit of MeV, and thus, one can extract the constraint on $L$ as $32 \le L \le 112$ MeV.

\begin{figure}[!htb]
\centering
\includegraphics[scale=0.5]{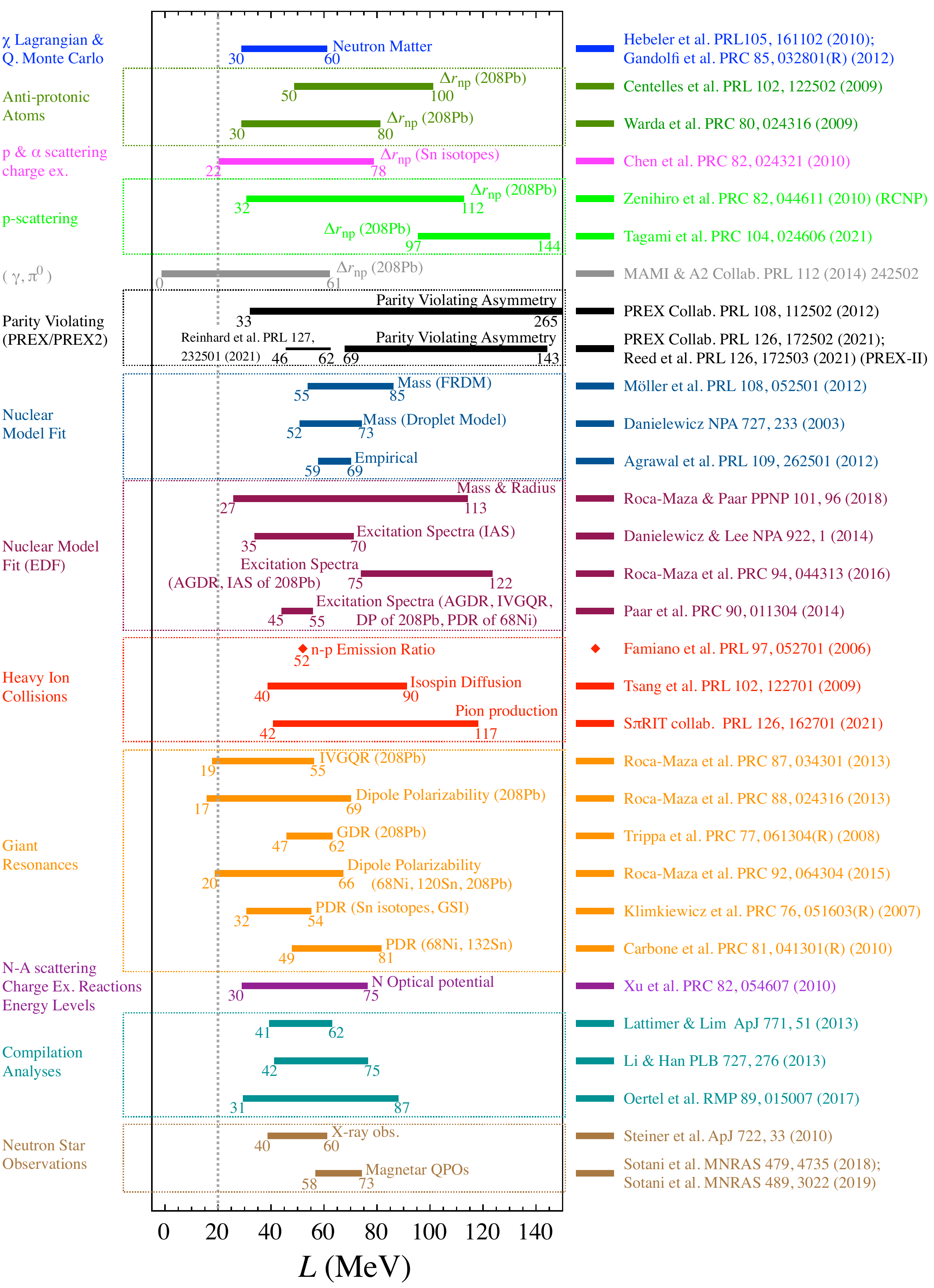} 
\caption{Constraints on $L$ from various experiments and astronomical observations, updated from the figure in Ref.~\cite{Vinas:2013hua}, using the data in Ref.~\cite{ROCAMAZA201896} together with the constraints on $L$ discussed in this study.
FRDM, IAS, AGDR, IVGQR, DP, PDR, GDR, and $\Delta r_{np}$ stand for
finite-range droplet model,
isobaric analog states,
anti-analog giant dipole resonance,
isovector giant quadrupole resonance,
dipole polarizability,
pygmy dipole resonance,
giant dipole resonance,
and 
neutron-skin thickness,
respectively. 
At the result with PREX-II, the constraint on $L$ with $L=54\pm 8\, \mathrm{MeV} $ derived by the reanalysis is also shown (see text for the detail.) 
The vertical dotted line denotes the lower limit of $L$ constrained from the condition that the pure neutron matter should not have a quasi-bound state \cite{Oyamatsu:2017qzv}.
\label{fig:const-L}}
\end{figure}

In Fig.~\ref{fig:const-L}, we show the constraint on $L$ obtained by nuclear experiments focused in this study together with the previous constrains by updating the figure shown in Ref. \cite{Vinas:2013hua}. In this figure, we show the value of $L$ constrained from 
the nucleon matter calculated with quantum Monte Carlo techniques \cite{PhysRevLett.105.161102,Gandolfi:2011xu}; 
the neutron-skin thickness for antiprotonic atoms \cite{PhysRevLett.102.122502,PhysRevC.80.024316}; 
the neutron-skin thickness in the $p$ and $\alpha$ scattering \cite{PhysRevC.82.024321}; 
the neutron-skin thickness in the $p$-scattering \cite{Zenihiro:2010zz,PhysRevC.104.024606}; 
the neutron-skin thickness in the $(\gamma,\pi^0)$ reaction \cite{PhysRevLett.112.242502}; 
the parity-violating asymmetry (PREX/PREX-II) \cite{PhysRevLett.108.112502,PhysRevLett.126.172502,Reed:2021nqk}; 
the empirical nuclear model fit \cite{PhysRevLett.108.052501,DANIELEWICZ2003233,PhysRevLett.109.262501}; 
the nuclear model fit thorough energy density functionals (EDF) \cite{DANIELEWICZ20141,PhysRevC.90.011304,PhysRevC.94.044313,ROCAMAZA201896}; 
heavy ion collisions \cite{PhysRevLett.97.052701,PhysRevLett.102.122701,SRIT:2021gcy}; 
nuclear giant resonances \cite{PhysRevC.76.051603,PhysRevC.77.061304,PhysRevC.81.041301,Roca-Maza:2012uor,Roca-Maza:2013mla,PhysRevC.92.064304}; 
nucleon optical potentials \cite{PhysRevC.82.054607}; 
compilation analyses \cite{Lattimer_2013,LI2013276,RevModPhys.89.015007}; 
and several NS observations \cite{Steiner_2010,Sotani:2018tdr,Sotani:2019pja}.
The vertical dotted line denotes the constraint on $L$ as $L\gsim 20$ MeV with the condition that
the pure neutron matter should not have a quasi-bound state \cite{Oyamatsu:2017qzv}.

\textit{4.~Constraints on the NS mass and radius relation.}
As mentioned before, any EOSs have their own values of $K_0$ and $L$, with which in turn each EOS is characterized. Nonetheless, it may be generally difficult to discuss the dependence of the NS mass and radius on two parameters. To solve this difficulty, an auxiliary parameter, $\eta$, is proposed \cite{Sotani:2013dga}, which is a combination of $K_0$ and $L$ given by 
\begin{equation}
  \eta = (K_0 L^2)^{1/3}. \label{eq:eta}
\end{equation}
We note that $\eta$ has been empirically found, where its physical meaning is still uncertain.
Thanks to the introduction of $\eta$, one can systematically discuss the NS mass and radius with one parameter, directly adopting the experimental data. In practice,
the mass $M$ and gravitational redshift $z\equiv (1-2GM/Rc^2)^{-1/2}-1$ for the low-mass NS with the radius $R$, whose central density $\rho_{\text{c}}$ is less than twice the nuclear saturation density, are expressed as functions of $u_{\text{c}}\equiv \rho_{\text{c}}/\rho_0$ and $\eta$, where $G$, $c$, and $\rho_0$ denote the gravitational constant, the speed of light, and the nuclear saturation density, such as
\begin{align}
  M/M_\odot &= 0.371 - 0.820 u_{\text{c}} + 0.279 u_{\text{c}}^2 -\left(0.593 -1.25 u_{\text{c}} + 0.235 u_{\text{c}}^2\right)\eta_{100},  \label{eq:mm}  \\
  z &= 0.00859 - 0.0619 u_{\text{c}} + 0.0255 u_{\text{c}}^2 - \left(0.0429 - 0.108 u_{\text{c}} + 0.0120 u_{\text{c}}^2\right)\eta_{100}, \label{eq:z}
\end{align}
where $\eta_{100}$ denotes $\eta/(100\ {\rm MeV})$ \cite{Sotani:2013dga}. Combining $M(u_{\text{c}},\eta_{100})$ and $z(u_{\text{c}},\eta_{100})$, one can plot the mass and radius for a given values of $u_{\text{c}}$ and $\eta$. 
We note that this technique is applicable only for the low-mass NSs, because the addtional effects, such as many-body effect and/or appearance of additional composition, should be taken into account, when the central density becomes somewhat large.  
\par
\begin{figure}[!htb]
\centering
\includegraphics[scale=0.7]{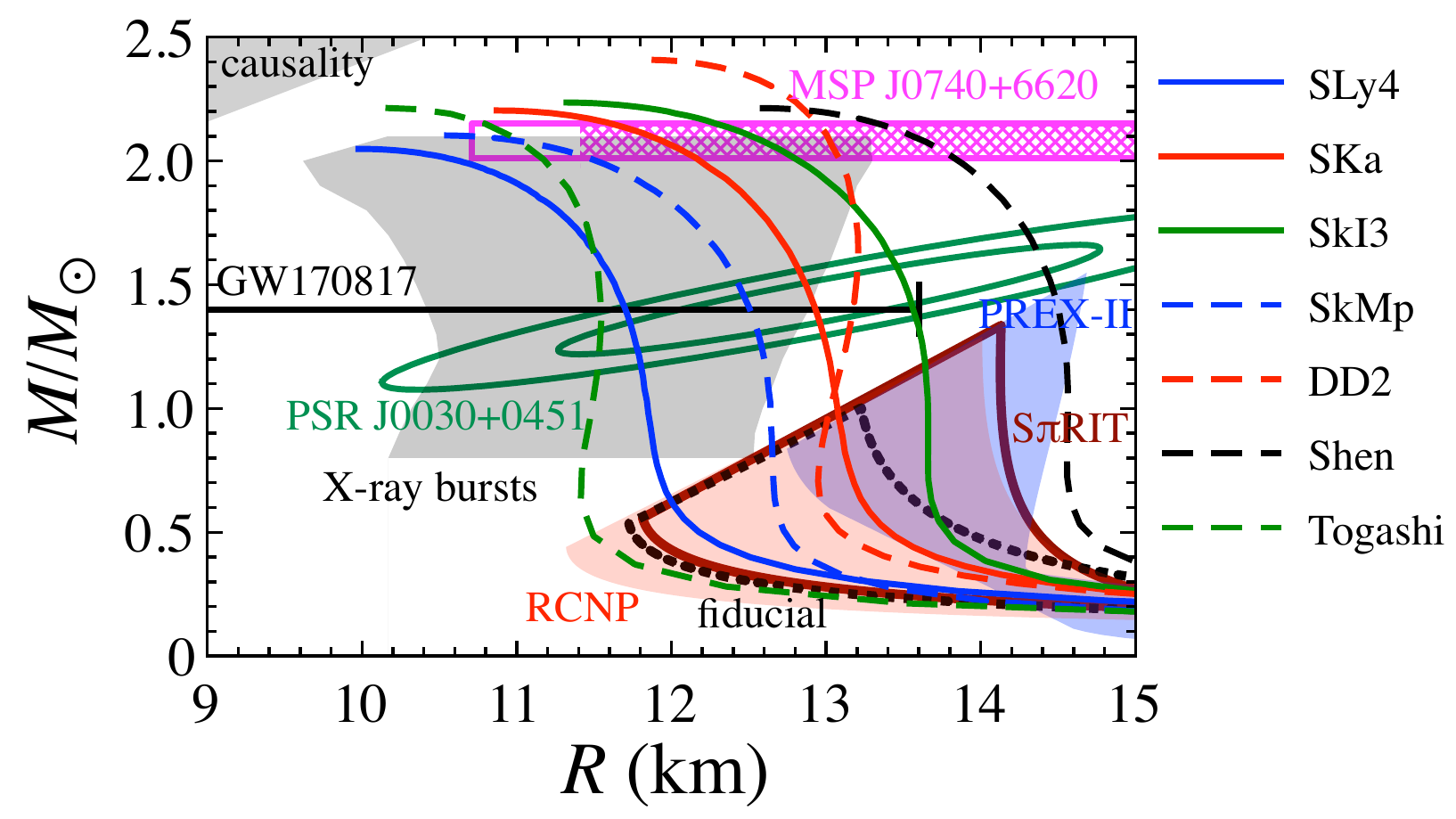} 
\caption{The constraints derived from the nuclear experiments are put on the bottom-right part, where the constraining region in the order from left to right corresponds to RCNP, S$\pi$RIT, and PREX-II.
For reference, the fiducial region is also shown, assuming that $L=60\pm 20 \, \mathrm{MeV}$ and $K_0=240\pm 20 \, \mathrm{MeV}$.
In addition, we show the astrophysical and theoretical constraints are shown (see text for details). For reference, NS mass and radius relations constructed with five different EOSs listed in Table \ref{tab:EOS} are also shown. 
The constraint from MSP J0740+6620 is shown by the shaded region ($68\%$) and the enclosed region with solid line ($95\%$).}
\label{fig:MR}
\end{figure}

In Fig.~\ref{fig:MR}, we show the resultant constraint on the NS mass and radius relation by adopting the constraints on $L$ 
obtained from the nuclear experiments discussed in the previous section, 
together with the constraint on $K_0$ as $K_0 = 240\pm20$ MeV, which correspond to $72.9\le \eta \le 152.7$ MeV for S$\pi$RIT, $101.6\le \eta \le174.5$ MeV for PREX-II, and $60.8\le \eta\le 147.9$ MeV for RCNP. 
For reference, we also show the NS mass and radius relation as a fiducial region by assuming the fiducial values of $L$ and $K_0$, i.e., $L=60\pm 20 \, \mathrm{MeV}$ and $K_0=240\pm 20 \, \mathrm{MeV}$, which corresponds to $70.6 \le\eta\le 118.5$ MeV. 
The constraints can be put on the bottom-right part in this figure, i.e., for the low density region, where we consider the NS model whose central density is up to twice the saturation density.
Note that the PREX-II reanalysis ($77.5 \le \eta \le 100.0 \, \mathrm{MeV} $) is consistent with the results obtained by RCNP and S$\pi$RIT,
and thus, hereinafter, it is not explicitly shown. 
One can eventually constrain the EOS for NS matter, because the NS mass and radius relation predicted by the EOS has to pass through the allowed region shown in Fig.~\ref{fig:MR}. 

In order to compare these constraints derived from the experiments, in Fig.~\ref{fig:MR} we show the four different constraints from astrophysical observations and one theoretical constraint. The constraint on the NS radius comes from the GW170817 \cite{Annala:2017llu}, i.e., the $1.4M_\odot$ NS radius should be less than 13.6 km, considering the tidal deformability observed in the gravitational waves from the binary NS merger. The NS with maximum mass observed so far is MSP J0740+6620 \cite{2021ApJ...915L..12F}, 
whose mass is $M/M_{\odot} = 2.08\pm 0.07$. 
The Neutron star Interior Composition Explorer (NICER) basically gives the constraint on the stellar compactness, $M/R$, on PSR J0030+0451 by carefully observing the pulsar light curve \cite{Miller:2019cac,Riley:2019yda}. The resultant constraint is shown by the tilted ellipses, where inner and outer edges correspond to the $1\sigma$ (68\%) and $2\sigma$ (95\%) constraints \cite{Blaschke:2020qqj}. 
NICER also gives us the radial constraint on PSR J0740+6620, i.e., $12.39_{-0.98}^{+1.30}$ km \cite{Riley_2021} and $13.7_{-1.5}^{+2.6}$ km \cite{Miller_2021}.
Through the X-ray burst observations from NSs, one can also constrain the NS mass and radius, as in Ref. \cite{Steiner:2012xt}, where for example the $1.4 M_\odot$ NS radius lies between $10.4$ and $12.9 \, \mathrm{km}$. Meanwhile, from the causality, one can exclude the top-left region corresponding to $R < 2.824GM/c^2$ \cite{Lattimer:2012nd}. By comparing the astrophysical and theoretical constraints mentioned here, we can say that all constraints derived from the nuclear experiments are still consistent. 

Moreover, in Fig.~\ref{fig:MR}, for reference, we also plot the mass and radius relations for the NS models constructed with several EOSs, i.e.,  the EOSs based on the Skyrme-type effective interaction, such as 
SLy4 \cite{CHABANAT1998231,Douchin:2001sv}, 
SKa \cite{KOHLER1976301}, 
SkI3 \cite{REINHARD1995467}, SkMp \cite{PhysRevC.40.2834},
the EOS based on the relativistic framework, such as 
DD2 \cite{PhysRevC.89.064321} and Shen \cite{Shen:1998gq}, and the EOS constructed with the variational many-body calculation, Togashi \cite{TOGASHI201778}. The EOS parameters and the maximum mass for the NS constructed with each EOS are listed in Table \ref{tab:EOS}.
The EOSs selected here except for the Shen satisfy the astronomical constraints shown in Fig.~\ref{fig:MR}, although SLy4 may be marginal to the constraint from MSP J0740+6620, and they are roughly consistent with the region constrained by three nuclear experiments. 
We note that, in the present study, the Shen is selected for reference because it has been adopted as a standard EOS, even though it has been ruled out from the constraint from GW170717.
Meanwhile, considering the allowed area given by the nuclear experiments (together with the astronomical restrictions), 
some of the EOSs may be ruled out. For example, the Togashi EOS passes through the lower $L$ boundary, 
which may be ruled out if the constraints from S$\pi$RIT and PREX-II are strictly true.
On the other hand, the Shen EOS is consistent only with the higher $L$ (the higher $M$ and $R$) covered by PREX-II, but it is inconsistent with some of the astronomical restrictions. 
Overall, the area covered by the S$\pi$RIT seems to agree with other constraints without significant inconsistencies.

\begin{table}
\caption{EOS parameters adopted in this study, $K_0$, $L$, and $\eta$, and the maximum mass, $M_{\rm max}$, for the NS constructed with each EOS.} 
\label{tab:EOS}
\centering
\begin{tabular}{l|cccc}
\hline\hline
EOS & \multicolumn{1}{c}{$K_0$ (MeV)} & \multicolumn{1}{c}{$L$ (MeV)} & \multicolumn{1}{c}{$\eta$ (MeV)} & \multicolumn{1}{c}{$M_{\rm max}/M_\odot$}   \\
\hline
SLy4
 & 230 & 45.9 &  78.5 & 2.05  \\ 
SKa
 & 263 & 74.6 & 114 & 2.22  \\ 
SkI3
 & 258 & 101 & 138 & 2.25  \\ 
SkMp
 & 231 & 70.3 & 105 & 2.11  \\ 
DD2
 & 243 & 55.0  & 90.2  & 2.41  \\ 
Shen
 & 281 & 111  & 151  &  2.17   \\  
Togashi
 & 245  & 38.7  & 71.6 & 2.21   \\ 
\hline \hline
\end{tabular}
\end{table}

{\it 5. Conclusion.}
Terrestrial nuclear experiments must be important for constraining the NS mass and radius relation especially for a low-density region, which can complement the constraint obtained from the astrophysical observations. In this study, we show which region in the NS mass and radius relation is allowed by using the recent constraints on the density-dependence of nuclear symmetry energy obtained via S$\pi$RIT and PREX-II together with the experiment by RCNP.
Compared to the other astrophysical constraints on the NS mass and radius, the 
allowed region we gave in this study, based on the nuclear experiments, still seems to be consistent, but the improvement of terrestrial experiments certainly helps us to understand the equation of state for NS matter. A number of future experiments are planned, which are expected to provide a further constraint on the NS mass and radius relation.
For example, CREX measurement with $^{48} \mathrm{Ca}$ has already been done, which may tell us the additional constraint.
In addition, the constraint on the higher-order saturation parameters, such as $K_{\text{sym}}$ and $Q_0$, is also important for constraining the NS EOS, e.g., \cite{PhysRevD.105.063010}.

\section*{Acknowledgment}
The authors are grateful to X.~Roca-Maza and X.~Vi\~{n}as for giving us some data of $L$ constraints in Fig.~\ref{fig:const-L} and to A.~Dohi for providing the data of the NICER constraint in Fig.~\ref{fig:MR}. They also acknowledge R.~Hirai, T.~Isobe, K.~Grande-Otsuki, D.~Suzuki, and T.~Takiwaki for fruitful discussion. This work is supported in part by JSPS KAKENHI (Grant Numbers:
JP19J20543,  
JP19KK0354,  
JP19H00693,  
JP20H04753,  
JP20H05648,  
JP21H01087,  
JP21H01088),  
by the RIKEN Pioneering Program for ``Evolution of Matter in the Universe (r-EMU)'' and the RIKEN Incentive Research Project.

%
\bibliographystyle{./ptephy}
\bibliography{sample631}
%












\end{document}